\documentclass[doublespacing]{elsart}
\usepackage{epsfig}
\usepackage{amssymb}
\begin{document}
\begin{frontmatter}
\title{A time dependent solution for the operation of ion chambers in a high 
ionization background.}
\author[Wisc]{Christos Velissaris\thanksref{now}}
\ead{christos@hep.wisc.edu}
\address[Wisc]{University of Wisconsin, 1150 University Avenue, Madison WI, 53706}
\thanks[now]{Address for Correspondence: 1150 University Avenue,University of Wisconsin, Madison WI, 53706}
\begin{abstract}
We have derived a time dependent solution describing the development of space 
charge inside an ion chamber, subjected to an externally caused ionization rate N. The solution enables the derivation of a formula that the operational parameters of the chamber must satisfy for saturation free operation. This formula contains a correction factor to account for the finite duration of the ionization rate N.
\end{abstract}
\end{frontmatter}

\section{Introduction}

It is a well known effect that the operation of ion chambers in a high 
ionization environment is limited by the space charge accumulated in the 
chamber active volume. This space charge is dominated by the positive ions 
since the electrons move with a drift velocity 1,000 times larger. 
The resulting electric field inside the chamber is no more uniform and 
it takes its least absolute value near the anode.\\
When this value reaches zero 
the electron collection stops and the chamber becomes saturated.
It is well known \cite{1,2,3,4} that a distinction between long and 
short pulses 
should be made when attempting to calculate saturation effects.
In the short pulse approximation the duration of ionization and charge 
collection time is short with respect to the time the positive ions need to 
move appreciably with respect to the distance d between the chamber electrodes.
In this approximation the positive ions are considered practically immovable 
throughout the 
whole electron collection time, thus the space charge density inside the 
chamber is uniform and constant. 
In the long pulse approximation the duration of ionization is long enough that 
a steady state is reached. In this state the number of collected positive 
ions per unit time is equal to the number of generated by the external 
ionization factor. The space chargg is no more depending on time but only 
on the position x between the chamber electrodes.\\
In this paper we derive a time dependent solution of the space charge 
accummulated in the chamber. We have made the assumption that the positive 
ions are moving with a constant drift velocity 
$V_{drift}$=$\mu\cdot\frac{V_{0}}{d}$ throughout the whole ionization and 
charge collection time.
This solution collapses into the short pulse approximation for ionization 
times much less than the characteristic time of the chamber $T_{0}$, and 
yields the steady state solution for ionization times larger or equal to 
$T_{0}$. We derive a simple expression for the quantity $T_{0}$, the 
characteristic time of the ion chamber, as well as for the electric fiels 
inside the chamber as a function of space and time.\\
Finally we derive a modified saturation equation,
which defines the conditions that the operational paramenters of the ion 
chamber should satisfy in order for the detector to operate free of saturation 
effects. This equation is a function not only of the positive ion mobility 
$\mu$, the 
electrode gap d and the externally applied voltage $V_{0}$, but a function of 
the duration T of the external ionization as well. \\  
Throughout the discussion we have assumed that the cathode plate is kept at 
potential -V$_{0}$ at x=0 and the anode is grounded at x=d. In our 
calculations we have used the esu system of units.

\section{The Steady State}

The accummulation of space charge inside the ion chamber, subjected in a 
constant external ionization rate N, is governed by the continuity equations:
\newline\newline
${\int_{x}^{d}}{{\frac{\partial{p(x^{\prime},t)}}{\partial{t}}}d{x^{\prime}}}=N(d-x)-p(x,t)u_{p}(x,t)$
for the positive ions, and:
\newline
${\int_{0}^{x}}{{\frac{\partial{n(x^{\prime},t)}}{\partial{t}}}d{x^{\prime}}}=Nx-n(x,t)u_{n}(x,t)$
for the electrons.\cite{1,4,9}
\newline\newline
Here u$_{p}$ is the positive ion and u$_{n}$ the electron drift velocities.
For x=d p(d,t)=0, and for x=0 n(0,t)=0 that is the consentration of positive ions next to the 
anode and the consentration of electrons 
next to the cathode are always zero. Also at t=0 p(x,0)=n(x,0)=0. 
In the steady state:\\
p(x)=$\frac{N(d-x)}{u_{p}(x)}$ and n(x)=$\frac{Nx}{u_{n}(x)}$ \\ 
Since the mobility of the electrons is 1,000 larger than the positive ions 
we can ignore the electron contribution to the spacecharge and thus we will 
write for the steady state:\\ \\
$\rho$(x)=$\frac{N(d-x)}{\mu{|E(x)|}}$ \\ \\
where E(x) is the electric field in 
the active volume of the chamber in the steady state and $\mu$ is the positive 
ion mobility.
By solving the Maxwell equations we get for the electric field:\\ \\
$|E(x)|=\sqrt{E(d)^{2}+\frac{4{\pi}N}{\mu}\cdot(d-x)^{2}}$
\newline \newline
If we imposing the initial conditions V(0)=-V$_{0}$ and V(d)=0 we get the 
relationship:\\ \\
${{{V_{0}}\over{d^2}}\cdot\sqrt{{\mu}\over{N\pi}}}=\sqrt{1+z^2}+z^{2}ln({\frac{1+\sqrt{1+z^2}}{z}})$ with z=$\sqrt{{{\mu}E(d)^{2}}\over{4{\pi}Nd^{2}}}$ \\ \\
For real values of z the second part of the equation is greater than 1, so in 
order for E(d) to exist: \\ \\
$\frac{V_{0}}{d^{2}}{\geq}\sqrt{{N\pi}\over{\mu}}$ \\ \\
This is a condition the operating parameters of the ion chamber must satisfy 
in order for the ion chamber to operate 
free of saturation effects, provided that the steady state has been reached.\\
If we assume that the positive ions are moving with constant drift velocity 
$\mu\cdot\frac{V_0}{d}$ then the steady state space charge and electric field 
become: \\ \\ 
$\rho(x)\approx\frac{Nd(d-x)}{\mu{V_0}}$ \\  
$E(x)\approx-\frac{V_0}{d}+\frac{2{\pi}Nd}{{\mu}{V_0}}(-\frac{2{d^2}}{3}-x^{2}+2dx)$. \\ \\
Under this approximation the steady state saturation condition yields: \\ \\
$\frac{V_{0}}{d^{2}}{\geq}\sqrt{{2{\pi}N}\over{3\mu}}$. \\ \\
This approximate condition based on the assumption of constant positive ion 
drift velocity differs from the one based on the exact positive ion drift 
velocity, by a correction factor $\sqrt{\frac{3}{2}}\approx$1.22.

\section{The Short Pulse Approximation}

When the ion chamber is subjected to an intense short lived ionization pulse 
the short pulse approximation is more suitable to describe the space 
charge effectsin the operation of the detector.
We can model the space charge density inside 
the chamber at the end of the ionization pulse as $\rho$, constant 
throughout the chamber volume.
Solution of Maxwell equations for the electric field yields:\\ \\
E(x)=4$\pi\rho${x}-2$\pi\rho${d}-$\frac{V_{0}}{d}$=-4$\pi$(x$_{sat}$-x)
where x$_{sat}$=$\frac{d}{2}$+$\frac{V_{0}}{4{\pi\rho}d}$ \\ \\
The electric field becomes zero at x=x$_{sat}$. Thus, in order for the chamber 
to work free of saturation effects $x_{sat}{\geq}d$ which leads to the 
condition: \\ \\
${\frac{V_{0}}{d^2}}{\geq}2\pi\rho$. \\ \\
An interesting consequence of the short pulse approximation is that always, a 
portion of the fast moving electrons is collected. If we suppose that the 
whole 
ionisation happens instantaneously the fast electron swarm will keep 
moving (and collected at the anode) from x=0 to x=x$_{sat}$ where the electric 
field becomes zero. We have considered the electron drift velocity as a 
product 
of the electron mobility times the Electric field. If we suppose that all the 
electrons 
from x=0 to x=x$_{sat}$ have been collected and electrons between x=x$_{sat}$ 
and x=d have been lost (since the drift velosity has become zero), the 
collection efficiency can be estimated as \\ \\
$\epsilon_{coll}=\frac{x_{sat}}{d}=\frac{1}{2}+\frac{V_0}{4\pi\rho{d^{2}}}$ 
if $x_{sat}\leq$d otherwise $\epsilon_{coll}$=1 \\ \\
We have ignored all recombination effects during the electron movement. 
We see that for extremely fast pulses the collection efficiency is at least 
0.5, even with the chamber operating under saturation conditions. The 
collection efficiency also increases linearly with the applied voltage V$_{0}$ 
and it is inverseley proportional to the charge density $\rho$. Recently 
conducted experiments have observed this effect.\cite{10,11}

\section{A Time Dependent Solution.}

\begin{figure}[!ht]
\begin{center}
\includegraphics[width=12cm]{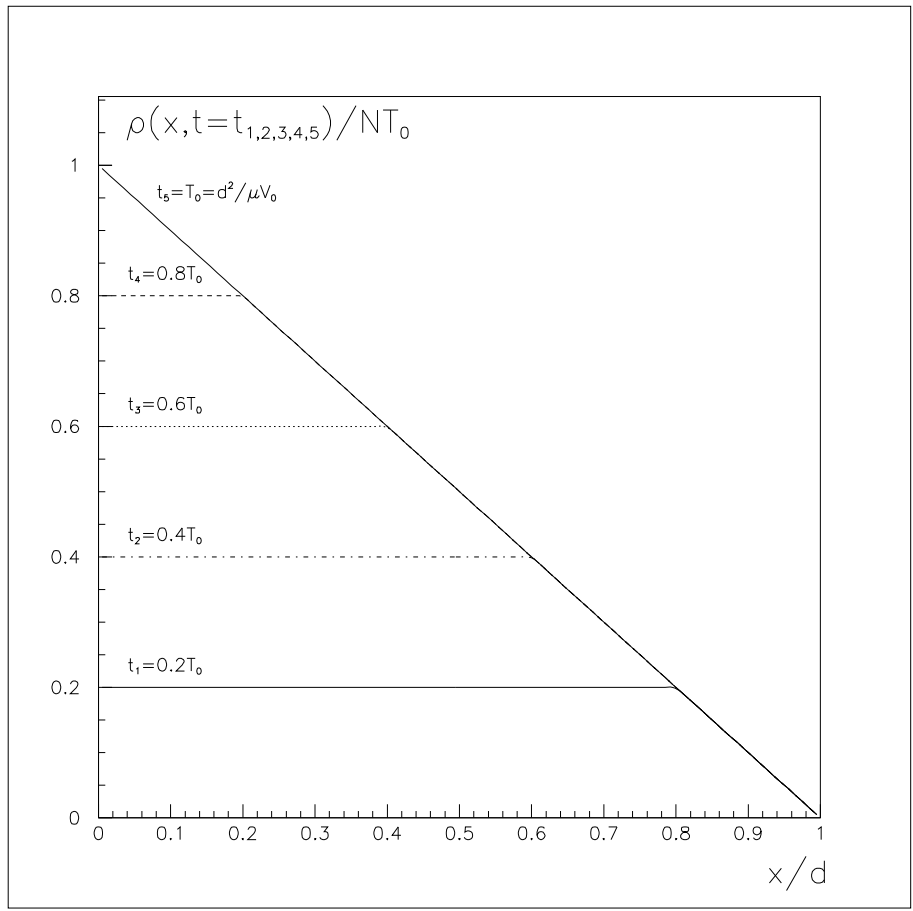}
\end{center}
\caption[capt1]{In this plot we present the space charge distribution inside the 
ion chamber for various times t$_{1}$,t$_{2}$,t$_{3},$t$_{4}$,t$_{5}$. At 
t=t$_{5}$ the steady state is reached and the space charge distribution 
becomes stationary thereafter.}
\label{figure1}
\end{figure}

\begin{figure}[!ht]
\begin{center}
\includegraphics[width=12cm]{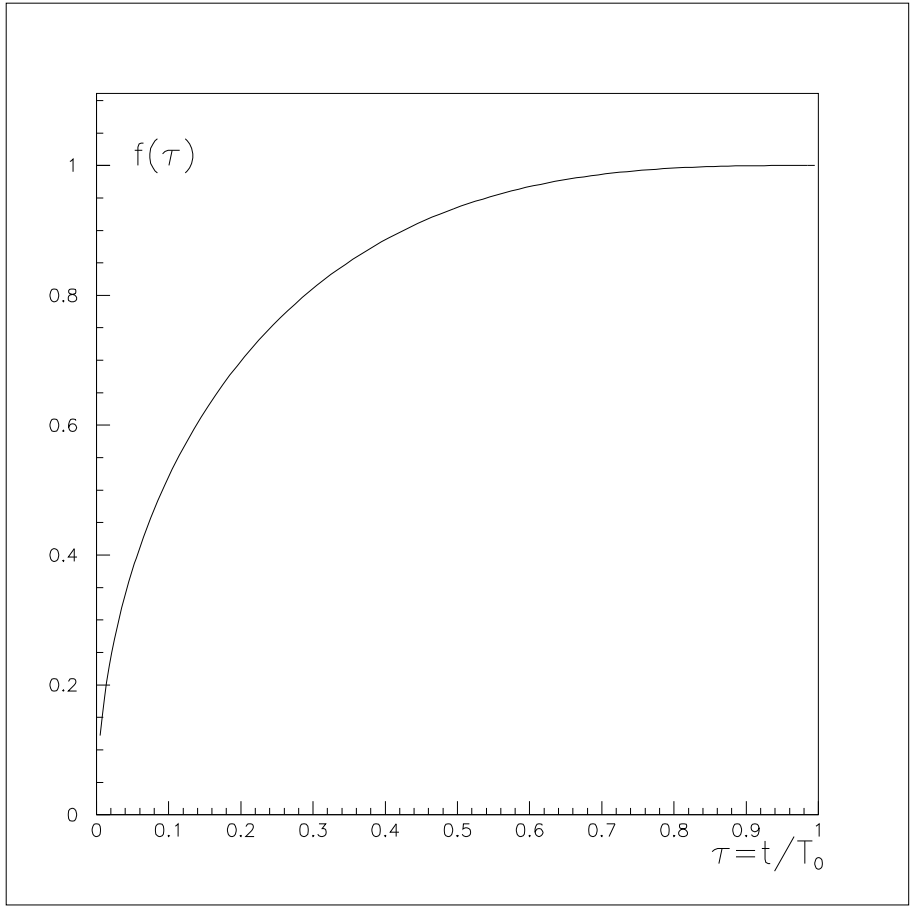}
\end{center}
\caption[capt2]{In this plot we present the lower limit 
$f(\tau)$=$\sqrt{{\frac{2{\pi}N}{3\mu}}\cdot{\tau(\tau^{2}-3\tau+3)}}$ of 
the quantity $\frac{V_{0}}{d^2}$ in order for an ion chamber to operate 
saturation free
as a function of the parameter $\tau$=$\frac{t}{T_0}$ for 0$\leq\tau\leq$1.
For simplicity we have taken $\frac{2{\pi}N}{3\mu}$=1.}
\label{figure2}
\end{figure}

The discussion below is based on the assumption that the positive ions are 
moving with constant drift velocity $\mu\cdot{{V_0}\over{d}}$ throughout the 
whole chamber volume and duration of charge collection. \\
We have derived a solution 
describing the time dependence of the accumulated space charge 
inside the chamber active volume. It is described by the equation: \\ \\
$\rho(x,t)=\frac{Nd}{{\mu}V_0}\cdot{x_0}$ when $0{\leq}x{\leq}d-x_{0}$ and \\ 
$\rho(x,t)=\frac{Nd}{{\mu}V_0}\cdot(d-x)$ when $d-x_{0}{\leq}x{\leq}d$ \\
with $x_0={{{\mu}{V_0}}\over{d}}\cdot{t}$ \\ \\
At time t the accumulated space charge is constant for values of x 
less than d-x$_{0}$ and a linear function of x for values greater than 
d-x$_{0}$, where $x_{0}= {{\mu{V_0}}\over{d}}\cdot{t}$. The steady state is 
reached when $x_{0}$=d, that is, when $t=T_{0}=\frac{d^{2}}{\mu{V_0}}$. 
We called $T_{0}$ the characteristic time of the ion chamber.\\  
In Figure \ref{figure1} we present the space charge distribution inside the 
ion chamber for various times.
By solving the Maxwell equations we can calculate the electric field inside 
the ion chamber as a function of x and t.\\ \\
$E(x,t)=E(d)-{{{2{\pi}Nd}\over{\mu{V_{0}}}}\cdot{x_{0}^{2}}}-
{{{4{\pi}Nd}\over{\mu{V_{0}}}}\cdot{x_0}(d-x_{0}-x)}$ for 
0${\leq}$x${\leq}$d-x$_{0}$  and \\
$E(x,t)=E(d)-\frac{2{\pi}Nd}{\mu{V_0}}\cdot{(d-x)^{2}}$ for 
d-x$_{0}{\leq}$x${\leq}$d \\ \\
E(d) is the electric field value at the anode at x=d. It can be calculated 
from the boundary values V(x=0,t)=-V$_{0}$ and V(x=d,t)=0 . We find: \\ \\
$E(d)=-{V_{0}\over{d}}+{{2{\pi}N}\over{{\mu}V_{0}}}\cdot{x_{0}}\cdot{(d^{2}-x_{0}d+{{x_{0}^{2}}\over{3}})}$ with $x_{0}= {{\mu{V_0}}\over{d}}\cdot{t}$ \\ \\
In order for the chamber to work free from saturation effects E(d)$\leq$0, 
so we arrive at the saturation condition: \\\\
${\frac{V_0}{d^2}}{\geq}\sqrt{{\frac{2{\pi}N}{3\mu}}\cdot
{\tau(\tau^{2}-3\tau+3)}}$ for $\tau{\leq}$1 and \\
${\frac{V_0}{d^2}}{\geq}\sqrt{\frac{2{\pi}N}{3\mu}}$ for $\tau{\geq}$1 \\ 
with $\tau={\frac{t}{T_0}}$ and $T_{0}={\frac{d^2}{\mu{V_0}}}$ the 
characteristic time of the ion chamber. In Figure \ref{figure2} we present 
the upper limit of the quantity $\frac{V_0}{d^2}$ as a function of the 
parameter $\tau$ for 0$\leq\tau\leq$1. For $\tau\geq$1 the steady state has 
been reached and this upper limit remains constant, independent of $\tau$. 
For simplicity we have assumed $\frac{2{\pi}N}{3\mu}$=1.\\ 
For a pulsed beam of duration T we have, N=${\rho}\over{T}$. Here, N is the 
ionization charge density rate and $\rho$ is the total charge density having 
been produced 
inside the ion chamber by the pulse after time T. We can then write: \\ \\
${{V_{0}}\over{d^2}}\geq\sqrt{{{2{\pi}\rho}\over{3{\mu}T_{0}}}\cdot({{T^{2}}\over{T_{0}^{2}}}-3{{T}\over{T_0}}+3)}$ for $T{\leq}T_{0}$ and \\
${{V_{0}}\over{d^2}}\geq\sqrt{{2{\pi}\rho}\over{3{\mu}T_{0}}}$ for 
$T{\geq}T_{0}$. \\ \\
For  $\tau{\geq}1$ ($T{\geq}T_{0}$) that is for times greater than the 
characteristic time of 
ion chamber T$_{0}$ we retrieve the steady state solution. For times t much 
less than T$_{0}$ if we approximate $\tau^{2}$-3$\tau$+3$\approx$3 we get:\\
${{V_{0}}\over{d^2}}{\geq}{2{\pi}Nt}$.\\
Since N is the ionization density rate in the chamber, Nt is the total charge 
density $\rho$ produced by the pulse after time t. We then retrieve 
the short pulse approximation condition: \\
${{V_{0}}\over{d^2}}{\geq}{2{\pi}\rho}$.

\section{Conclusion}

After a brief review of the operation of ion chambers in an intense ionization 
environment we presented a formula describing the time development of the 
space 
charge inside the chamber. The formula collapses into the short pulse 
approximation for short enough ionization pulses with respect to the 
characteristic time of chamber T$_{0}$ and reproduces the steady state 
solution if the externally caused ionization lasts long enough.
Although the formula has been derived under the assumption that the positive 
ions are moving always with a constant drift velocity $\mu\cdot{V_0}\over{d}$, 
it helps us to understand (qualitatively and up to some degree quantitatively)
 the operation of ion chambers. More specifically we deduce:
\begin{itemize}

\item In order to account for the finite duration of an externally caused 
ionization density rate N in an ion chamber we have derived the following
 modified saturation condition: \\ \\
${\frac{V_0}{d^2}}{\geq}\sqrt{{\frac{2{\pi}N}{3\mu}}\cdot
{\tau(\tau^{2}-3\tau+3)}}$ for $\tau{\leq}$1 and
${\frac{V_0}{d^2}}{\geq}\sqrt{\frac{2{\pi}N}{3\mu}}$ for $\tau{\geq}$1 \\
with $\tau={\frac{t}{T_0}}$ and $T_{0}={\frac{d^2}{\mu{V_0}}}$ the 
characteristic time of the ion chamber. \\ \\
For a pulsed beam of duration T, N=${\rho}\over{T}$ where $\rho$ is the total
charge density produced in the chamber by the 
pulse after time T. We can then write: \\ \\
${{V_{0}}\over{d^2}}\geq\sqrt{{{2{\pi}\rho}\over{3{\mu}T_{0}}}\cdot({{T^{2}}\over{T_{0}^{2}}}-3{{T}\over{T_0}}+3)}$ for $T{\leq}T_{0}$ and
${{V_{0}}\over{d^2}}\geq\sqrt{{2{\pi}\rho}\over{3{\mu}T_{0}}}$ for 
$T{\geq}T_{0}$.\\

\item Both, the short pulse approximation and the steady state represent valid 
solutions in judging saturation effects in ion chambers. 

\item The steady state is reached throughout the whole chamber volume 
within a finite time T$_{0}$ (the characteristic time of the ion chamber) 
after the onset of the external ionization process. However, as we showed in 
this paper, different points reach the steady state at different times.

\item For short enough pulsed beams with respect to the ion chamber 
characteristic time the positive ion mobility $\mu$ does not play any role 
in judging saturation effects. \cite{10}

\item Saturation effects should be judged by taking into account not only the 
chamber operating voltage, gap and ion mobility but the duration of the 
ionization process as well.

\item Whether we assume that the positive ions move with constant drift 
velocity or their speed is affected by the variation of the 
electric field due to the space charge, for short pulses both solutions 
collapse into the short pulse approximation and no distinction is made. 
Corrections may be important when the steady state is formed due to the 
different space distribution of the positive ion charge density inside the 
chamber. 
The maximum correction must be applied when steady state is reached. As we 
have seen that factor is $\sqrt{\frac{3}{2}}\approx$1.22.   
An effort to solve the continuity equations, that describe the time 
development of the positive ion distribution in the chamber, by using 
computational techniques is presented in a recent paper\cite{8}. 

\end{itemize}


\begin{thebibliography}{99}
\bibitem{1} J. Sharpe, ``Nuclear radiation Detectors'', John Wiley.
\bibitem{2} J. W. Boag ``Ionization Measurements at very high Intensities''.
I. Pulsed Radiation Beams. Brit. J. Radiol. 23, 601 (1950)
\bibitem{3} J. W. Boag ``The Saturation Curve for Ionization Measurements 
in Pulsed radiation Beams'' Brit. J. Radiol. 25, 649 (1952)
\bibitem{4} J. W. Boag ``Space Charge Distortion of the Electric Field in a 
Plane parellel  Ionization Chamber'' Phys. Med. Biol. 8, 461 (1963)
\bibitem{5} J. W. Boag and T. Wilson, ``The Saturation curve at High 
Ionization Density'' Brit. J. Appl. Phys., vol. 3, pp. 222-229 (1952)
\bibitem{6} J. W. Boag in F. H. Attix, W.C. Roesch ``Radiation Dosimetry'', 
Academic Press.
\bibitem{7} ``Ionization Dosimetry at High Intensity''. In proceedings of the 
International School of Physics E. Fermi, Course XXX, ``Radiation Dosimetry'', 
p. 70.
\bibitem{8} S. Palestini, G. D. Barr, C. Biino, P.O. Calafiura, A. Ceccucci, 
C.Cerri et. al., ``Space Charge in Ionisation Detectors and the NA48 
Electromagnetic Calorimeter''  Nuclear Instruments and Methods in Physics 
Research A421 (1999) pp. 75-89.
\bibitem{9} C. Velissaris. ``Principles of ionization chamber operation under 
intense ionization rates.'' NuMI-NOTE-BEAM-0717, unpublished
\bibitem{10} J. McDonald, C. Velissaris, B. Viren, M. Diwan, A. Erwin, 
D. Naples, H. Ping ``Ionization chambers for monitoring at high intensity 
charged particle beams.'' Nucl. Instrum. Meth. A496:293-304, 2003
\bibitem{11} R. Zwaska et.al.``Beam Tests of Ionization chambers for the 
NuMI neutrino beam'' IEEE Trans. Nucl. Sci.50:1129-1135, 2003

\end{thebibliography}
\end{document}